\documentclass[manuscript,screen,nonacm=true]{acmart}
\usepackage{fontspec}

\usepackage{booktabs}
\usepackage{graphicx}
\usepackage{hyperref}
\usepackage{natbib}
\usepackage{url}

\begin{document}

\title[Domus]{Domus: An Open-Data Web Platform for House History Research}

\author{David M. Straub}
\orcid{0000-0001-5762-7339}
\email{david.straub@hm.edu}
\affiliation{%
  \institution{Munich University of Applied Sciences HM}
  \streetaddress{Dachauer Str. 98b}
  \city{Munich}
  \postcode{80335}
  \country{Germany}
}
\correspondingauthor

\begin{abstract}
House history -- the record of who lived in a building, who owned it, and how its address changed over time -- is a central concern of genealogical and local history research. Wikidata and OpenHistoricalMap together provide an open-data infrastructure, CC0-licensed and capable of representing this data: Wikidata through its linked-data property model with temporal qualifiers, OpenHistoricalMap through historical building footprints with start and end dates. What has been missing is a domain-specific tool that makes contributing and discovering building history accessible to genealogists and local historians. Domus fills this role: a map-based web platform that guides users through structured, crowdsourced editing of building records in Wikidata and displays corresponding OpenHistoricalMap footprints. The application requires no custom backend -- authentication, data reading, and data writing all operate through public APIs in the browser, and all data are stored under CC0 in public collaborative databases. This paper describes the genealogical data model, the architecture, and key technical design decisions.
\end{abstract}


\maketitle

\section{Introduction}\label{sec:intro}

Building history sits at the intersection of genealogy and local history: understanding where an ancestor lived, who their neighbors were, and how a street evolved over decades requires linking persons to specific buildings over time. This is a long-recognized research need -- the \textit{Häuserbuch} tradition in German-speaking countries, which documents building occupancy house by house, dates back centuries -- yet the corresponding data infrastructure remains largely pre-digital and geographically fragmented. The knowledge at stake -- who lived next to whom, which trades occupied a street's ground floors, when a house burned down or was rebuilt -- can be assembled from parish registers, directories, and municipal archives, but the assembled result typically remains in a researcher's private notes and rarely outlives its compiler.

Genealogical software such as Gramps supports the representation of buildings and places alongside family records; its web-based counterpart Gramps Web extends this with integration of OpenHistoricalMap and other historical map layers, combining person data with geographic and temporal context for research. Such tools cannot, however, provide an open repository for structured building history -- address timelines, occupant and ownership records, construction and demolition dates -- with long-term archiving and reliable sourcing. Such data, where it exists at all, is held in local \textit{Häuserbuch} projects, heritage registries, or municipal archive portals: typically not interoperable, and rarely openly licensed for reuse. Regional projects illustrate this landscape: \textit{Haus und Hof} \cite{hausundhof}, run by the Austrian Society for Research on Genealogy and Regional Heritage (ÖFR), offers a map-based interface to house history data for Lower Austria; FRANZI \cite{franzi}, an Innsbruck University initiative, georeferences historical cadastral survey sheets for Tyrol and Vorarlberg. FRANZI's data is openly licensed, while \textit{Haus und Hof} is not; both, however, remain tied to their own regional interface rather than a shared, general-purpose infrastructure. What is missing is a map-based, genealogically oriented interface built directly on top of such a general-purpose, open data infrastructure -- not bounded by region or institution.

Two open collaborative databases together provide the infrastructure needed for such a platform. Wikidata \cite{vrandevcic2014wikidata} offers a linked-data model that accommodates arbitrary temporal qualifiers on statements, enabling the representation of address history, occupant records, and ownership changes with precise date ranges; its CC0 license ensures contributed data remain permanently open. OpenHistoricalMap (OHM) \cite{openhistoricalmap} contributes what Wikidata lacks: building shapes and other historical geographic features such as roads, rivers, and administrative boundaries, modeled as elements carrying \texttt{start\_date} and \texttt{end\_date} tags. OHM elements can carry a \texttt{wikidata=*} tag identifying the corresponding Wikidata item, providing a cross-reference between the two databases. Together they form a coherent open substrate for house history that neither database alone provides.

Domus is built on both foundations. It presents a map-based interface for discovering buildings recorded in Wikidata, displaying their genealogical record alongside OHM footprints, and contributing new data through guided editing forms designed for genealogical use cases. It also writes the \texttt{wikidata=*} cross-reference back to OHM when a user links a building, authenticated through a separate OHM account. The application requires no custom backend -- all reading and writing operates through public APIs in the browser. Domus is a project of CompGen (Verein für Computergenealogie e.V.); the code is open source under the MIT license at \href{https://github.com/compgen-ev/domus}{https://github.com/compgen-ev/domus}, deployed as a static site via GitHub Pages and served under CompGen's own domain at \href{https://domus.genealogy.net}{https://domus.genealogy.net}. Giving house history a shared, open, map-based home in the Wikidata/OHM infrastructure lets it accumulate as a crowdsourced dataset that any local historian, family researcher, or resident can query or extend.

\section{Data Model}\label{sec:datamodel}

The genealogical requirements for a building record are: its type and name, address history with temporal ranges, persons who occupied or owned it and when, construction and demolition dates, heritage status, links to geographic representations, and -- critically for genealogical use -- the documented source of every assertion.

These requirements map onto a well-defined subset of Wikidata properties (Table~\ref{tbl:properties}). Address history is represented with P6375 (street address), a free-text string qualified with P580/P582 (start/end time). Free text was chosen over structured address items because it handles the full variety of historical address forms naturally: modern street-and-number addresses, pre-grid village house numbers (\textit{Nr. 42}), and addresses that changed following administrative reorganizations. Occupants (P466) and owners (P127) are linked person items, also qualified with P580/P582. Construction and demolition use P571 and P576 respectively. All dates require an explicit calendar-model URI in Wikidata's data model; following the Wikidata convention, Domus automatically selects the proleptic Julian calendar for years before 1583 and the Gregorian calendar thereafter.\footnote{A deliberate simplification: the actual transition varied by region, with some Protestant German territories retaining the Julian calendar until 1700. The calendar model of any date can be corrected in Wikidata directly where a source demands it.} Beyond exact day/month/year values, Domus accepts and displays dates at decade, century, and millennium precision, and represents genuinely unresolved dates as an unknown-value statement bracketed by P1319 (earliest date) and/or P1326 (latest date) qualifiers -- e.g. ``before 1409'' or ``between 1380 and 1409'' -- matching the level of precision that pre-modern archival sources typically support. Heritage designation (P1435) is displayed as a badge but not editable through Domus, to avoid duplicating the curation workflows of heritage authorities. Where a building has been demolished and rebuilt, or otherwise succeeded, Domus displays the P167/P1398 (replaced by / replaces) link to the successor or predecessor item, letting users follow a site's continuity across reconstructions.

\begin{table}
\centering
\caption[]{Wikidata properties used by Domus.}
\label{tbl:properties}
\begin{tabular}{p{\dimexpr 0.500\linewidth-2\tabcolsep}p{\dimexpr 0.500\linewidth-2\tabcolsep}}
\toprule
Property & Role \\
\hline
P31 (instance of) & Building type \\
P625 (coordinate location) & Geographic position \\
P571 / P576 (inception / dissolved) & Construction and demolition dates \\
P1319 / P1326 (earliest / latest date) & Date-uncertainty qualifiers \\
P6375 (street address) & Address history (+ P580/P582 qualifiers) \\
P466 (occupant) & Historical residents (+ P580/P582) \\
P127 (owned by) & Historical owners (+ P580/P582) \\
P1435 (heritage designation) & Protected status (display only) \\
P167 / P1398 (replaced by / replaces) & Building succession \\
P854 / P4656 / P485 / P248 & Source reference (URL, archive, or book) \\
\bottomrule
\end{tabular}
\end{table}

A key design constraint is mandatory sourcing: Domus hard-blocks form submission unless the user provides a source reference, reflecting the genealogical norm that every assertion must cite its origin. Mandatory sourcing also aligns Domus with Wikidata's notability policy, which accepts items referring to a clearly identifiable entity that can be described using serious and publicly available references -- a criterion a sourced building record satisfies by construction. Person items created for occupants and owners additionally fulfill a structural need in the sense of that policy, making the statements on building items more useful, and are in practice backed by the same archival sources. Three source types are supported: an online source (reference URL via P854, or P4656 if the URL points to a Wikimedia project such as Wikipedia or Wikisource, with optional page number via P304); an archival document (the holding archive as a Wikidata item via P485, a call number via P217, and optional page via P304); and a book, either as an existing Wikidata item (P248, stated in) or as a free-text entry supplying title (P1476), author (P2093), and year (P577). All statements created or modified in a single editing session receive the same reference block. For physical archival records, contributors can additionally document their findings on CompGen's GenWiki\footnote{\href{https://wiki.genealogy.net}{https://wiki.genealogy.net}} and cite the resulting page as an online source, making the evidence verifiable without a visit to the archive.

OHM models a building as one or more way (or relation) footprints, each carrying \texttt{start\_date} and \texttt{end\_date} tags for the period it was valid; a building that was rebuilt over time is represented as a sequence of such elements rather than a single static footprint. Any of these elements can carry a \texttt{wikidata=*} tag cross-referencing the corresponding Wikidata item. Domus fetches all footprints tagged with a building's Wikidata QID via the Overpass API \cite{overpassapi}, renders them on the map with a time slider for stepping through successive periods, and links back to OHM for further editing.

\section{Architecture}\label{sec:architecture}

\begin{figure}[!htbp]
\centering
\includegraphics[width=0.8\linewidth]{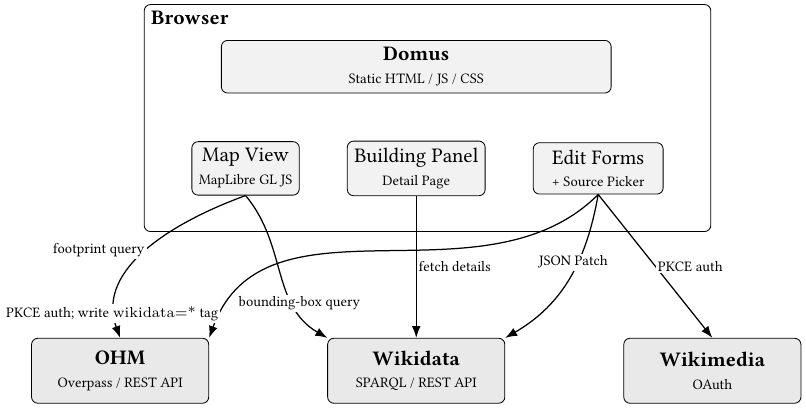}
\caption[]{Architecture of Domus. All data flows through public APIs; no custom backend server is required.}
\label{fig:architecture}
\end{figure}

Domus is a static single-page application: the production artifact is a bundle of HTML, CSS, and JavaScript served as static files with no server-side processing. Figure~\ref{fig:architecture} shows the system structure.

\begin{figure}[!htbp]
\centering
\includegraphics[width=0.95\linewidth]{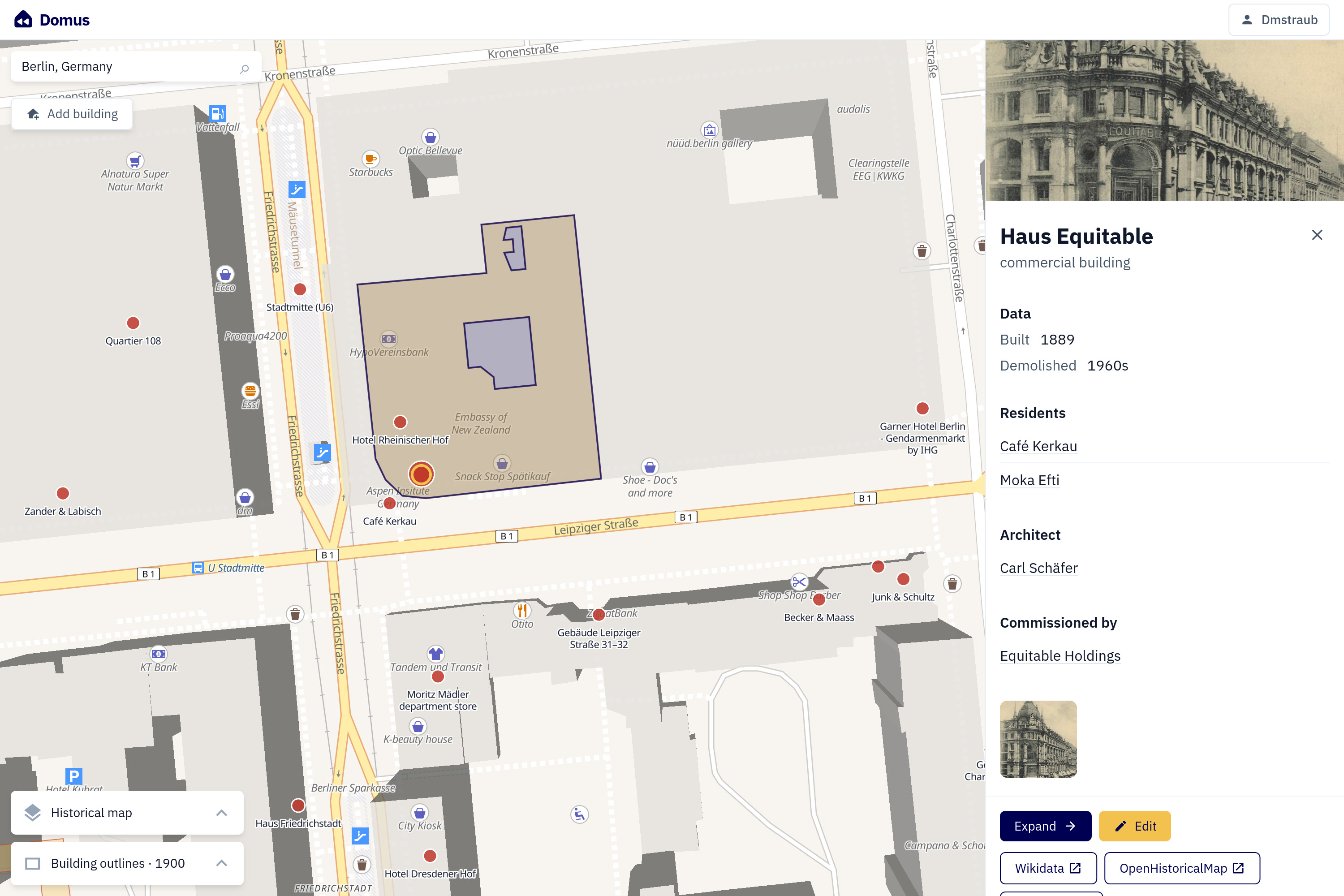}
\caption[]{Domus showing a selected building with its side panel open. The map displays Wikidata building pins; the OHM footprint of the selected building is rendered as an overlay.}
\label{fig:screenshot}
\end{figure}

The frontend is built with Lit 3 \cite{lit} web components in TypeScript, bundled with Vite 6 \cite{vite}. The map layer uses MapLibre GL JS \cite{maplibregljs} with OpenFreeMap \cite{openfreemap} tiles as the base. The interface is fully responsive: on desktop, selecting a building slides in a side panel (Figure~\ref{fig:screenshot}); on mobile, a bottom drawer opens over the map. A geolocation button allows users to center the map on their current position, making it practical to add or annotate buildings on-site from a mobile device. A full detail page at \texttt{?id=Q\dots} provides the complete building record on any screen size.

Map-driven discovery uses bounding-box SPARQL queries issued on map movement (triggered at zoom level $\geq$ 14). The query fetches all entities with a coordinate (P625) within the viewport, along with their P31 type, optional image (P18), and construction date (P571). A typical urban viewport returns 1,000--3,000 entities from the Wikidata SPARQL endpoint in 700--800 ms. Filtering these to buildings requires checking membership in the P279 subclass hierarchy rooted at the generic building class (extended with P31 instances of ``type of building'', Q811102), which is five or more levels deep with well over 9,000 distinct types. Server-side SPARQL property-path traversal of this hierarchy causes timeouts when combined with a spatial filter on thousands of entities; instead, Domus pre-computes the complete set of building-type identifiers offline and ships it as a static asset of well under 200 KB, reducing membership checks to under 1 ms per entity at query time.

Authentication uses OAuth 2.0 \cite{rfc6749} with PKCE (Proof Key for Code Exchange) \cite{rfc7636} entirely in the browser, run independently against Wikimedia and against OHM, since writing the \texttt{wikidata=*} cross-reference back to an OHM element requires its own OHM account authorization. PKCE avoids the need for a client secret: the client proves possession of a one-time code verifier at token exchange, so an authorization code intercepted in the front channel cannot be redeemed by an attacker; a state parameter guards against CSRF during the callback. No backend token exchange is required for either flow. Access tokens are stored in \texttt{localStorage}, which is accessible to any JavaScript running on the page. This is an accepted trade-off for a platform whose protected resources -- the Wikidata REST API and the OHM API -- are themselves public services with a full audit trail. The tokens grant write access under the authenticated user's own account, not to a private server: a compromised token would permit edits attributable to the victim's account, but no access to private data, and all edits remain permanently attributable and reversible. The application stores no personal data beyond these tokens; the only other local state is edit timestamps, and all contributed content is immediately public under CC0.

Editing uses the Wikidata REST API via JSON Patch \cite{rfc6902} operations: individual statements are added to or modified on existing items without overwriting unrelated data. Entity search (for person pickers in occupant and owner fields) uses the Wikidata \texttt{wbsearchentities} action with debounced input. All Wikidata endpoints -- SPARQL, REST API, and entity search -- support cross-origin requests, which is a prerequisite for the frontend-only architecture: without CORS headers the browser's same-origin policy would block direct API access from the application's domain. Because the REST API and SPARQL endpoint operate on separate database replicas with a replication lag of typically 5--60 seconds, Domus tracks recent edit timestamps in \texttt{localStorage} and displays a staleness notice if the SPARQL result has not yet caught up, scheduling background re-fetches until it does. Because all writes go directly to Wikidata and OHM, Domus also inherits their community quality-control infrastructure: every edit appears in public recent-changes feeds and watchlists and can be reverted with the standard tools of either project.

The application is internationalized into German and English using \texttt{@lit/localize}, with runtime locale detection from \texttt{navigator.languages}. German is supported alongside English as CompGen is a German association, but the framework makes adding further languages straightforward.

\section{Status and Outlook}\label{sec:status}

Domus is in early-stage production at \href{https://domus.genealogy.net}{https://domus.genealogy.net}. Currently implemented are:

\begin{itemize}
\item geographic map browsing with building discovery, including pin clustering at low zoom;
\item a full building detail view with a photo gallery drawing on P18 images;
\item edit forms for labels, aliases, address history, occupant and owner history, construction and demolition dates, architect, and commissioned-by;
\item OHM footprint display with a time slider for navigating building shapes across historical periods;
\item historical raster map overlays as selectable base-map layers, e.g. 19th- and early-20th-century cadastral and topographic surveys from regional archives such as LGL-BW, swisstopo, and the Staatsbibliothek zu Berlin, each included only because it carries an open license; the layer registry is a small, self-contained configuration, so adding further regions is straightforward wherever a suitably licensed historical map source exists;
\item OAuth 2.0 login against both Wikimedia and OHM;
\item building creation from a map pin, with optional pre-fill from OHM data;
\item writing the bidirectional \texttt{wikidata=*} cross-reference back to OHM elements.
\end{itemize}

Wikidata pins are intentionally shown without date filtering, reflecting that the point-in-time state of a building record is less meaningful than the full timeline displayed in the detail view. Where a Wikidata item already carries a GOV identifier (P2503), Domus surfaces it as an outbound link to the corresponding GOV (Genealogisches Ortsverzeichnis)\footnote{\href{https://gov.genealogy.net}{https://gov.genealogy.net}} entry. Deeper GOV integration remains limited, since GOV's building coverage is sparse and skewed toward administrative and heritage-scale objects rather than ordinary houses.

Some limitations follow directly from the architectural choices. The frontend-only design makes Domus dependent on the availability and rate limits of public endpoints -- the Wikidata SPARQL and REST services and the Overpass API -- with no caching layer of its own to absorb outages or throttling. For the SPARQL queries, QLever \cite{qlever}, an alternative engine serving a public Wikidata mirror, has been suggested as a potentially faster backend than the Wikidata Query Service; it has not yet been evaluated in depth and is currently not used. And like any crowdsourcing tool, map-driven discovery inherits its starting conditions from the underlying database -- with uneven results. In many areas, Wikidata's existing building coverage is remarkably rich: heritage registers, church and monument inventories, and past import projects mean the map is well populated from day one, and Domus starts as a discovery tool rather than an empty canvas. Other regions, by contrast, hold few or no building items, and there browsing only becomes rewarding once the first contributors have created records; seeding from existing open datasets is an option worth exploring.

Another possible future improvement is a connection to the \textit{historische Adressbücher}\footnote{\href{https://adressbuecher.genealogy.net}{https://adressbuecher.genealogy.net}}, historical city directories digitized by CompGen volunteers, whose entries record occupants per address and year -- a natural fit for the data model, though unexplored so far.

\section{Conclusions}\label{sec:conclusions}

Domus demonstrates that a genealogically useful house history platform can be built without a custom backend, by treating Wikidata and OpenHistoricalMap as the authoritative data layer rather than as secondary imports or mirrors. Every contribution made through Domus enriches the global open knowledge graph immediately and is queryable by any other Wikidata client. The frontend-only architecture eliminates backend hosting costs and operational complexity for a volunteer-run community initiative. More broadly, the combination of Wikidata's linked-data model and OHM's temporal geographic layer provides a reusable infrastructure pattern for digital humanities projects that need to represent historical spatial data alongside person records -- a recurring need in genealogy, local history, and historical prosopography. What changes in practice is who can contribute a street's history: not a single institution, but any descendant, neighbor, or local archivist with relevant knowledge, with the result immediately visible to anyone else viewing the same map.

\section*{AI-Assisted Development}
Domus's implementation was developed with substantial assistance from Claude (Anthropic), an AI coding assistant used interactively throughout development in 2026. The author conceived the project, defined the data model (Data~Model), and made the architectural and design decisions described in this paper -- including the choice of OAuth 2.0 with PKCE for dual Wikimedia/OHM authentication, the JSON Patch editing strategy, the offline precomputation of building-type identifiers, and the replication-lag handling between the Wikidata SPARQL and REST endpoints -- and directed Claude in implementing them across the frontend components, API integrations, and build tooling. Claude also generated the TypeScript type annotations and the unit test suite covering the utility functions, under the author's direction. The author reviewed all AI-generated code and tests, ran static type-checking across the entire TypeScript codebase and the unit test suite, and validated application behavior against the live Wikidata and OpenHistoricalMap APIs, iterating with Claude on any issues found; the author is responsible for the integrity of the reported work.

\begin{acks}
The author thanks the CompGen community (Verein für Computergenealogie e.V.) for providing the subdomain, domain expertise on genealogical data requirements, and beta testing; Minh Nguyễn for advice on Wikidata that was crucial to assess the feasibility of the project; Christopher Ernestus, Hermann Hartenthaler, Jesper Zedlitz, Georg Fertig, Petra Hildebrandt, Florian Straub, and Benjamin Klein for discussion that shaped several of the features described here; the Wikidata community for maintaining the open knowledge infrastructure on which this platform depends; OpenStreetMap US for running OpenHistoricalMap; and Zsolt Ero for providing OpenFreeMap map tiles free of charge.
\end{acks}

\bibliographystyle{ACM-Reference-Format}
\bibliography{main.bib}

@misc{hausundhof,
	author = {{Österreichische Gesellschaft für Familien- und regionalgeschichtliche Forschung (ÖFR)}},
	year = {2026},
	note = {Accessed: 2026-07-13},
	title = {Haus und {Hof}},
	url = {https://huh.oefr.at/},
	howpublished = {https://huh.oefr.at/},
}

@misc{franzi,
	author = {{Universität Innsbruck, Institut für Geschichtswissenschaften}},
	year = {2026},
	note = {Accessed: 2026-07-13},
	title = {FRANZI: Der {Franziszeische} {Kataster} f{\" u}r {Tirol} und {Vorarlberg}},
	url = {https://geschichte-franzi.uibk.ac.at/franzi/},
	howpublished = {https://geschichte-franzi.uibk.ac.at/franzi/},
}

@article{vrandevcic2014wikidata,
	author = {Vrande{\v c}i{\' c}, Denny and Kr{\" o}tzsch, Markus},
	journal = {Communications of the ACM},
	number = {10},
	year = {2014},
	pages = {78--85},
	publisher = {ACM New York, NY, USA},
	title = {Wikidata: a free collaborative knowledgebase},
	volume = {57},
}

@misc{openhistoricalmap,
	author = {{OpenHistoricalMap Contributors}},
	year = {2026},
	note = {Accessed: 2026-07-08},
	publisher = {OpenStreetMap U.S.},
	title = {OpenHistoricalMap},
	url = {https://www.openhistoricalmap.org},
}

@misc{overpassapi,
	author = {{OpenStreetMap Wiki Contributors}},
	year = {2026},
	note = {Accessed: 2026-07-08},
	title = {Overpass {API}},
	url = {https://wiki.openstreetmap.org/wiki/Overpass_API},
	howpublished = {https://wiki.openstreetmap.org/wiki/Overpass\textunderscore{}API},
}

@misc{lit,
	author = {{Google LLC and Lit Contributors}},
	year = {2026},
	note = {Accessed: 2026-07-08},
	title = {Lit},
	url = {https://lit.dev/},
	howpublished = {https://lit.dev/},
}

@misc{vite,
	author = {{VoidZero Inc. and Vite Contributors}},
	year = {2026},
	note = {Accessed: 2026-07-08},
	title = {Vite},
	url = {https://v6.vite.dev/},
	howpublished = {https://v6.vite.dev/},
}

@misc{maplibregljs,
	author = {{MapLibre Contributors}},
	year = {2026},
	note = {Accessed: 2026-07-08},
	title = {MapLibre {GL} {JS}},
	url = {https://maplibre.org/projects/gl-js/},
	howpublished = {https://maplibre.org/projects/gl-js/},
}

@misc{openfreemap,
	author = {{OpenFreeMap}},
	year = {2026},
	note = {Accessed: 2026-07-08},
	title = {OpenFreeMap},
	url = {https://openfreemap.org/},
	howpublished = {https://openfreemap.org/},
}

@misc{rfc6749,
	author = {Hardt, D.},
	doi = {10.17487/RFC6749},
	year = {2012},
	month = {10},
	publisher = {RFC 6749, RFC Editor},
	title = {The {OAuth} 2.0 {Authorization} {Framework}},
	url = {https://www.rfc-editor.org/rfc/rfc6749},
}

@misc{rfc7636,
	author = {Sakimura, N. and Bradley, J. and Agarwal, N.},
	doi = {10.17487/RFC7636},
	year = {2015},
	month = {9},
	publisher = {RFC 7636, RFC Editor},
	title = {Proof {Key} for {Code} {Exchange} by {OAuth} {Public} {Clients}},
	url = {https://www.rfc-editor.org/rfc/rfc7636},
}

@misc{rfc6902,
	author = {Bryan, P. and Nottingham, M.},
	doi = {10.17487/RFC6902},
	year = {2013},
	month = {4},
	publisher = {RFC 6902, RFC Editor},
	title = {JavaScript {Object} {Notation} ({JSON}) {Patch}},
	url = {https://www.rfc-editor.org/rfc/rfc6902},
}

@inproceedings{qlever,
	address = {New York, NY, USA},
	author = {Bast, Hannah and Buchhold, Bj{\" o}rn},
	booktitle = {Proceedings of the 2017 {ACM} on {Conference} on {Information} and {Knowledge} {Management} ({CIKM})},
	doi = {10.1145/3132847.3132921},
	year = {2017},
	pages = {647--656},
	organization = {ACM},
	title = {QLever: A {Query} {Engine} for {Efficient} {SPARQL}+{Text} {Search}},
}

\end{document}